\documentclass[aps,prb,twocolumn,superscriptaddress,showpacs,amssymb,amsmath,floatfix]{revtex4-1}

\usepackage{graphicx,color}
\usepackage{color,verbatim}
\usepackage{hyperref}

\newcommand{\bl}{\ensuremath{\lambda}}

\newcommand{\e}{\ensuremath{\varepsilon}}

\newcommand{\g}{\ensuremath{\mathbf{g}}}

\newcommand{\Ga}{\ensuremath{\Gamma}}
\newcommand{\Ha}{\ensuremath{H}}

\newcommand{\M}{\ensuremath{\mathbf{M}}}

\newcommand{\Om}{\ensuremath{\Omega}}

\newcommand{\T}{\ensuremath{\mathcal{T}}}

\begin{document}

\title{Nonequilibrium phonon backaction on the current noise in atomic-sized junctions}

\author{Tom\'{a}\v{s} Novotn\'y}
\email{tno@karlov.mff.cuni.cz}
\affiliation{Department of Condensed Matter Physics, Faculty of
Mathematics and Physics, Charles University in Prague, Ke Karlovu 5,
CZ-12116 Praha 2, Czech Republic}
\affiliation{Institut N{\'E}EL, CNRS and Universit{\'e} Joseph Fourier, BP 166, F-38042 Grenoble Cedex 9, France}

\author{Federica Haupt}\email{haupt@physik.rwth-aachen.de}
\affiliation{Institut f\"ur Theorie der statistischen Physik, RWTH Aachen University, D-52056 Aachen, Germany}
\affiliation{Fachbereich Physik, Universit\"at Konstanz, D-78457
Konstanz, Germany}

\author{Wolfgang Belzig}
\affiliation{Fachbereich Physik, Universit\"at Konstanz, D-78457
Konstanz, Germany}

\begin{abstract}
We study backaction effects of phonon heating due to tunneling electrons on the current noise in atomic-sized junctions. Deriving a generalized kinetic approximation within the extended Keldysh Green's functions technique, we demonstrate the existence of a characteristic backaction contribution to the noise in case of low external phonon damping. We provide a physically intuitive interpretation of this contribution at large voltage in terms of slow fluctuations of the phonon occupation, and show that it generally gives a significant correction to the noise above the phonon emission threshold.           
\end{abstract}

\date{\today}
 \maketitle


{\em Introduction.}
{Inelastic transport spectroscopy is an important tool of investigation for a wide range of phenomena  in nanojunctions, ranging from vibrations in atomic wires 
\cite{Agrait:PhysRep} to magnetic excitations of aggregates on surfaces.\cite{Eigler,*Heinrich} Inelastic studies provide both direct information on the non-electronic degrees of freedom, such as vibrations and/or spin excitations and, indirectly, additional knowledge of the electronic subsystem. Nowadays, experimental interest is not only limited to the mean current but it also extends to the current noise,\cite{Blanter} which provides further information on the transport properties of the junction, such as the number and transparency of the transmission eigenchannels.\cite{Djukic,*DiCarlo,*Kumar:platinum}  Inelastic effects on the current noise are also presently under experimental investigation,\cite{Kumar,*Scheer} accompanied by an  intense theoretical activity.\cite{Zhu,*Galperin:PRB06,Schmidt,Avriller,Haupt,Urban,HauptPRB10}

Since the electron-phonon ($e$-ph) interaction is often very weak, a perturbative approach to the lowest-order in $e$-ph coupling is usually appropriate for evaluating both the current\cite{Paulsson:RapCom,Viljas,delaVega} and noise,\cite{Schmidt,Avriller,Haupt,HauptPRB10} enabling semi-analytical treatments that can be combined with {\em ab initio} calculations.\cite{Frederiksen:PRB07,HauptPRB10}  There is, however, a tricky aspect of the problem, related to the heating of the vibrational mode(s) due to tunneling electrons, whose experimental signature is the voltage dependence of the conductance above the phonon emission threshold.\cite{Frederiksen:PRL04} Accounting for this effect pushes the theory well beyond the plain lowest order perturbation expansion. 

Nonequilibrium phonon heating has been taken into account either phenomenologically in terms of a rate equation for the phonon occupation,\cite{Frederiksen:PRL04, Koch:PRB06} or microscopically by infinite resummation of the electron-hole polarization bubble. \cite{Mitra,Viljas}  As far as only the current is concerned, the two approaches yield essentially identical results in the limit of weak $e$-ph coupling. However, they differ significantly in the case of the current noise,\cite{Urban} due to the feedback of the phonon dynamics on the statistics of the transmitted electrons, which is not captured by the phenomenological rate equation approach. According to Ref.~\onlinecite{Urban}, this feedback results in a cubic increase of the noise as a function of voltage at large bias, $S\sim V^{3}$. However, this prediction contradicts a preliminary unpublished  work by Jouravlev supervised by Nazarov\cite{Jouravlev} suggesting $S\sim V^{4}$. A clear physical interpretation of the results was not offered in either case.

Thus, despite its relevance for several systems of experimental interest,\cite{Agrait:PhysRep,Paulsson:RapCom,Ward:NatNano11} the effect of phonon heating on the current noise still remains an open problem.  In this work, we develop a consistent microscopic theory to address this issue. We derive an analytic kinetic-like expression for the phonon Green's function (GF), which  encodes the effects of electronic fluctuations and allows one to take into account the backaction of nonequilibrium phonons on all current cumulants. Focusing on the current noise, we demonstrate that current-driven fluctuations of the phonon occupation lead to a distinct correction to the noise related to the variance of the phonon occupation. We provide an intuitive physical interpretation of this correction, which at large voltage agrees qualitatively with the asymptotic result of Ref.~\onlinecite{Jouravlev}, i.e., $S\sim V^{4}$, and show that it generally leads to a significant increase of the noise above the phonon emission threshold in case of negligible phonon damping. Our predictions are relevant for currently ongoing experiments.\cite{Kumar,*Scheer}

{\em Model \& methods.} 
We consider the generic Anderson-Holstein type of Hamiltonian for  inelastic transport
through a nanojunction $\hat{\Ha}=\hat{H}_C+\hat{H}_{L,R}+\hat{H}_T$ with the central part $\hat{H}_C =\hat{H}_{0}+\hat{H}_{\rm ph}+\hat{H }_{e\rm ph},\,
\hat{H}_{0}=\sum_{i,j}h_{ij}d^{\dag}_{i}d_j,\, \hat{H}_{\rm ph} =\Om\hat{b}^{\dag}\hat{b},\,\hat{H}_{e\rm ph} =\sum_{i,j}M_{ij}d^{\dag}_id_j(b^{\dag}+b)$ and noninteracting electronic leads described by  $\hat{H}_{L,R}=\sum_{k,\alpha=L,R}\e_{\alpha,k}\hat{c}_{\alpha,k}^{\dag}\hat{c}_{\alpha,k}$, tunnel-coupled by $\hat{H}_T=\sum_{i,k,\alpha=L,R}(V_{\alpha,k}^{i}\hat{c}_{\alpha,k}^{\dag}\hat{d}_{i}+h.c.)$ to the central part (``dot''). The states of the leads are occupied according to the Fermi distribution $f_{\alpha}(\e)=(e^{\beta(\e-\mu_{\alpha})}+1)^{-1}$, with $\beta=1/k_{B}T$ and $\mu_{\alpha}$ being the chemical potential of lead $\alpha$. The applied bias voltage is $\mu_{L}-\mu_{R}=eV$. 
We consider here only a single vibronic mode (``phonon'') with energy $\Om$ --- other phonon modes simply contribute additively to the noise correction calculated below unless the phonon resonances overlap, in which case a matrix generalization of our approach would be needed.

To evaluate the noise, we employ the generalized Keldysh GF technique.\cite{GogolinKomnik,Nazarov:book} The key idea is to determine the GF of the dot $G_{\bl}^{ij}(t,t')=-i\hbar^{-1}\langle \mathcal{T}_{c}\hat{d}_{i}(t)\hat{d}_{j}(t')^{\dag}\rangle_{\bl}$ in the presence of a counting field $\lambda$, introduced as a time-dependent fictitious parameter in one of the tunneling matrix elements,\cite{GogolinKomnik} e.g., $V_{L,k}^{i}\to V_{L,k}^{i}e^{-\bl(t)/2}$ and ${V_{L,k}^{i\,*}}\to {V_{L,k}^{i\,*}}e^{\bl(t)/2}$, with $\bl(t)=\pm\bl \in\mathbb{R}$ on the forward (labeled by `$-$')/backward (`$+$') branch of the Keldysh contour.\footnote{For convenience, we consider here rotated counting field with respect to Ref.~\onlinecite{GogolinKomnik} $i\lambda\rightarrow\lambda\in\mathbb{R}$ to simplify the derivation of the generalized kinetic approximation by avoiding unnecessary complex factors.} Knowing $G_{\bl}^{ij}$, the current and the noise (and in principle all the other current cumulants) can be directly evaluated as described in detail in Refs.~\onlinecite{GogolinKomnik,HauptPRB10}. 
In previous studies,\cite{Haupt,HauptPRB10,Schmidt,Avriller} coupling to an external heat bath bringing phonons to equilibrium has been implicitly assumed. Here, we focus instead on the limit of {\em no external phonon damping}, where all the thermalization comes exclusively from the electronic degrees of freedom and strong nonequilibrium phonon heating effects are expected.  
 

{\em Generalized kinetic approximation.} 
To rigorously address the problem of nonequilibrium phonon heating, and its consequences for the transport properties of the junction, one needs to include the influence of the electrons on the phonons by dressing the phononic GF with the polarization operator $\check{\Pi}_{\lambda}$. To the lowest order in $e$-ph coupling, this  is given by the polarization bubble~\cite{Viljas,Mitra,Urban}
\begin{equation}\label{polarop}
\Pi^{\sigma\rho}_{\bl}(\e)=-i\int\frac{d \e'}{2 \pi}\mathrm{Tr}[\mathbf{M}\,\mathbf{g}^{\sigma\rho}_{\bl}(\e+\e')\,\mathbf{M}\,\mathbf{g}^{\rho\sigma}_{\bl}(\e')],
\end{equation}
with $\g_{\bl}^{\sigma\rho}\,(\sigma,\rho=\mp)$ being the Keldysh components of the dot's GF in the presence of the leads and of the counting field, but without $e$-ph coupling:\cite{GogolinKomnik,HauptPRB10}  
\begin{widetext}
\begin{equation}\label{electrGF}
\check{\mathbf{g}}_{\bl}(\e)=\begin{pmatrix}\e\mathbf{1} -\mathbf{h} -  i\sum_{\alpha=L,R}\mathbf{\Ga}_{\alpha}[f_{\alpha}(\e)-1/2]&   i\mathbf{\Ga}_Le^{\bl}f_L(\e)+i \mathbf{\Ga}_R f_R(\e)\\-i \mathbf{\Ga}_Le^{-\bl}[1-f_L(\e)]-i\mathbf{\Ga}_R[1-f_R(\e)]& -\e\mathbf{1}+\mathbf{h}-i\sum_{\alpha=L,R}\mathbf{\Ga}_{\alpha}[f_{\alpha}(\e)-1/2]\end{pmatrix}^{-1}.
\end{equation}
Here, the sign $\check{}$ stands for $2\times2$ matrices in Keldysh space and boldface letters indicate matrices in the dot electronic space indexed by the single-particle labels $i$ and $j$, e.g., $\mathbf{\Gamma}_{\alpha}=\{\Gamma^{ij}_{\alpha}\}$, with $\Gamma^{ij}_{\alpha}(\e)=2\pi \sum_{k}V_{\alpha,k}^{i}V_{\alpha,k}^{j^{*}}\delta(\e -\e_{k,\alpha})$ being the broadening due to coupling to lead $\alpha$.
Note that $\check{\Pi}_{\lambda}$ is explicitly $\lambda$-{dependent} and so is consequently the dressed phonon GF $\check{D}_{\bl}$ given by the Dyson equation: 
\begin{equation}\label{Dysoneq}
\check{D}_{\bl}(\e)=\begin{pmatrix}\frac{\e^{2}-\Om^{2}}{2 \Om}-\Pi_{\bl}^{--} & \Pi_{\bl}^{-+}\\
\Pi_{\bl}^{+-}&-\frac{\e^{2}-\Om^{2}}{2\Om}-\Pi_{\bl}^{++}
\end{pmatrix}^{-1}=
\frac{2\Om}{\mathcal{D}_{\bl}(\e)}\begin{pmatrix}\e^{2}-\Om^{2}+2\Om\Pi_{\bl}^{++} & 2\Om \Pi_{\bl}^{-+}\\
2 \Om\Pi_{\bl}^{+-}&-\e^{2}+\Om^{2}+2\Om\Pi_{\bl}^{--}
\end{pmatrix},
\end{equation}
with $\mathcal{D}_{\bl}(\e)=\big(\e^{2}-\Om^{2}+2\Om\Pi_{\bl}^{++}\big) \big(\e^{2}-\Om^{2}-2\Om\Pi_{\bl}^{--}\big)+4 \Om^{2}\Pi^{-+}_{\bl}\Pi^{+-}_{\bl} $. 
\end{widetext}

In the limit of weak $e$-ph coupling $1/{\mathcal{D}_{\bl}(\e)}$ shows two resonances around $\e\sim \pm \Om$. Close to these resonances,  $\mathcal{D}_{\bl}(\e)$ is given by $\mathcal{D}_{\bl}(\e\sim\pm\Om) \approx 4 \Om^{2}\big[(\e\mp\Om\pm\Delta_{\bl}(\pm\Om))^{2}+\xi_{\bl}(\pm\Om)/4 \big]$, with $\Delta_{\bl}(\e)=[\Pi^{++}_{\bl}(\e)-\Pi^{--}_{\bl}(\e)]/2$ and $\xi_{\bl}(\e)=4 \Pi^{-+}_{\bl}(\e)\Pi^{+-}_{\bl}(\e)-(\Pi^{++}_{\bl}(\e)+\Pi^{--}_{\bl}(\e))^{2}$. 
Neglecting the small  real frequency shift $\Delta_{\bl}(\e)$ and approximating in the distributional sense $\frac{x+a}{x^{2}+\Gamma^{2}/4}\stackrel{\Gamma \to 0}{\longrightarrow} \mathcal{P}\frac{1}{x}+ 2\pi \frac{a}{\sqrt{\Gamma^{2}}}\delta(x)$, where $\mathcal{P}$ stands for the principal value and $\delta(x)$ is the Dirac $\delta$-function, we arrive at the kinetic-limit expression -- i.e., with zero phonon line-width -- for the dressed phonon GF to the lowest (zeroth) order in $e$-ph coupling $\M$: 
\begin{equation}\label{Dgen}
\check{D}_{\bl}(\e)\!=\!\begin{pmatrix}
\mathcal{P}\frac{2 \Om}{\e^{2}-\Om^{2}}& 0\\
0 &-\mathcal{P}\frac{2 \Om}{\e^{2}-\Om^{2}} 
\end{pmatrix} \!-\!2\pi i\!\! \sum_{s=\pm}\delta(\e+s\Om)\check{\mathcal{N}}_{\bl}(\e), 
\end{equation}
where
\begin{equation*}
\check{\mathcal{N}}_{\bl}(\e)\!=\! \frac{i}{\sqrt{\xi_{\bl}(\e)}}\begin{pmatrix}
\frac{\Pi^{--}_{\bl}(\e)+\Pi^{++}_{\bl}(\e)}{2}& \Pi^{-+}_{\bl}(\e)\\
 \Pi^{+-}_{\bl}(\e) & \frac{\Pi^{--}_{\bl}(\e)+\Pi^{++}_{\bl}(\e)}{2}
\end{pmatrix},
\end{equation*}
$\check{\mathcal{N}}_{\bl}(-\e)=[\check{\mathcal{N}}_{\bl}(\e)]^{T}$, is the $\bl$-dependent generalized phonon occupation.  At $\bl=0$, it reduces to $\check{\mathcal{N}}_{\bl=0}(\e)=\left(\begin{smallmatrix}\bar{N}(\e)+1/2& \bar{N}(\e)\\ \bar{N}(\e)+1 & \bar{N}(\e)+1/2
\end{smallmatrix}\right)$, where $\bar{N}(\e)\equiv i{\Pi^{-+}_{\bl=0}(\e)}/{2|{\rm Im}\Pi^{R}(\e)|}$,   with $\Pi^{R}(\e)\equiv\Pi^{--}_{\bl=0}(\e)-\Pi^{-+}_{\bl=0}(\e)$,
is the {\it nonequilibrium} phonon occupation number, in agreement with Ref.~\onlinecite{Viljas}. In the kinetic limit, $\bar{N}\equiv \bar{N}(\Om)$ coincides with the phenomenological results of  Refs.~\onlinecite{Paulsson:RapCom,Frederiksen:PRB07}. It can be shown analytically that approximation \eqref{Dgen} preserves charge conservation for all cumulants to the lowest (second) order in the $e$-ph coupling.


 {\em Correction to noise due to nonequilibrium phonons.} 
The effects of phonon heating on the current noise $S$ can be
evaluated within the formalism of Ref.~\cite{HauptPRB10}, by replacing the free-phonon GF considered there with the dressed one, Eq.~\eqref{Dgen}.  Doing so, we can split the correction to the noise at the lowest order in the $e$-ph coupling in two parts $S_{e\rm ph}=S_{\rm av}+S_{\rm ba}$.  Here, $S_{\rm av}$ is the contribution due to coupling to a phonon  with  fixed average occupation, and it is simply given by the result of Ref.~\onlinecite{HauptPRB10} with the thermal phonon occupation replaced by the nonequilibrium one, $n_{B}(\Om)\to\bar{N}$.  In case of a single-level junction, it corresponds  to the phenomenological result of Ref.~\onlinecite{Haupt}. 
The additional term $S_{\rm ba}$ represents the  {\em backaction} of current-driven nonequilibrium fluctuations of the phononic occupation on the current noise itself and it reads   
\begin{equation}
S_{\rm ba}=-\frac{e^{2}}{\hbar}\sum_{\sigma,\rho=\mp}\int \frac{d\e}{2\pi}{\rm Tr}\Big\{\frac{\partial \g_{\bl}^{\sigma\rho}}{\partial \bl}\mathbf{\Xi}_{\bl}^{\rho\sigma} \Big\}_{\bl=0}
\end{equation}
with
\begin{equation}
\mathbf{\Xi}_{\bl}^{\sigma\rho}(\e)=i\int\frac{d\e'}{2\pi}\frac{\partial D^{\sigma\rho}_{\bl}(\e')}{\partial \bl}\M\, \g^{\sigma\rho}_{\bl}(\e-\e')\M.
\end{equation}
After some algebra, $S_{\rm ba}$ can be rewritten as
\begin{equation}\label{Noisecorr}
\begin{split}
\frac{S_{\rm ba}}{e^{2}/\hbar}&=-\frac{\Pi'_{++}+\Pi'_{--}-\Pi'_{+-}-\Pi'_{-+}}{|\mathrm{Im}\Pi^{R}(\Om)|}\\ 
&\times \Big\{\bar{N}(\bar{N}+1)(\Pi'_{++}+\Pi'_{--}-\Pi'_{+-}-\Pi'_{-+})\\
&+\big(\bar{N}+\tfrac{1}{2}\big)(\Pi'_{+-}-\Pi'_{-+})-\tfrac{1}{2}(\Pi'_{+-}+\Pi'_{-+})\Big\},
\end{split}
\end{equation}   
with $\Pi'_{\sigma\rho}\equiv \frac{\partial}{\partial \bl} \Pi^{\sigma\rho}_{\bl}(\Om)\big|_{\bl=0}$. This expression is valid for an {\em arbitrary} junction with weak $e$-ph coupling and represents one of the main results of this paper.\footnote{Derivation leading to  Eq.~\eqref{Noisecorr} can be easily generalized to account for partial phonon equilibration by a weakly coupled external heat bath provided the appropriate $\lambda$-independent phonon-self-energy term is added to Eq.~\eqref{polarop}. For details, see the supplement.\cite{Note3}}

Further progress can be made within the extended wide-band approximation (eWBA),\cite{Paulsson:RapCom,HauptPRB10} where an explicit expression for $S_{\rm ba}$ as a function of voltage and temperature can be derived. The full multilevel result is given in the supplemental material,\footnote{See enclosed Supplemental Material.} while here we discuss for simplicity only the paradigmatic case of a single-level junction symmetrically coupled to leads at $T=0$.\cite{Schmidt,Haupt} In this case we obtain the following constituents of Eq.~\eqref{Noisecorr}: $\bar{N}=\theta(|eV|-\Om)(|eV|/\Om-1)/4$, $\Pi'_{++}+\Pi'_{--}-\Pi'_{+-}-\Pi'_{-+}=i\gamma_{e\mathrm{ph}}(3-4\mathcal{T})eV/\pi$, $\Pi'_{+-}+\Pi'_{-+}=-i\gamma_{e\mathrm{ph}}(1-2\mathcal{T})eV/\pi$, $\Pi'_{+-}-\Pi'_{-+}=-i\gamma_{e\mathrm{ph}}(1-2\mathcal{T})\mathrm{sign}(V)\min(|eV|,\Om)/\pi$, and $|\mathrm{Im}\Pi^{R}(\Om)|=\gamma_{e\mathrm{ph}}\Om/\pi$, with $\gamma_{e \rm ph}=M^{2}\mathcal{T}^{2}/\Gamma^{2}$ being the dimensionless $e$-ph coupling constant and $\T\in[0,1]$ the elastic transmission coefficient.

{\em Analysis and discussion.} 
We note first  that $S_{\rm ba}$ is a strictly nonequilibrium correction, i.e., it is zero at $V=0$. This is consistent with the fluctuation-dissipation theorem, since there is no correction related to current-driven phonon fluctuations in the linear conductance. 
In the opposite limit of large bias voltage, $S_{\rm ba}$ is dominated by the term proportional to $\bar{N}(\bar{N}+1)$. Indeed this contribution grows like $V^{4}$ and $V^{3}$ for $eV>\Om$, while the remaining terms of  Eq.~\eqref{Noisecorr}, as well as the ``static'' term $S_{\rm av}$,\cite{Haupt} increase at maximum as $V^{2}$. 
The large-voltage behavior of $S_{e\rm ph}$ is then given by the leading terms of $S_{\rm ba}$
\begin{equation}\label{Sasym}
S_{e\rm ph}(eV\gg\Om)\approx\tfrac{e^{2}\gamma_{e\mathrm{ph}}}{\pi\hbar\Om} (3-4\mathcal{T})^{2}(eV)^{2}\bar{N}(\bar{N}+1).
\end{equation} 

This result can be fully understood in terms of a semiclassical mechanism related to slow fluctuations of the phonon occupation. The fluctuating occupation number $N(t)$ of a weakly driven oscillator is described by the master equation $dP_{n}(t)/dt=\gamma_{\downarrow}[(n+1)P_{n+1}(t)-nP_{n}(t)]+\gamma_{\uparrow}[nP_{n-1}(t)-(n+1)P_{n}(t)]$ for the probabilities $P_{n}(t)$ that $N(t)$ attains the value $n$ at time $t$.\cite{Zoller} The nonequilibrium rates can be microscopically evaluated as $\gamma_{\uparrow}=i\Pi^{-+}_{\bl=0}(\Om)/\hbar, \gamma_{\downarrow}=i\Pi^{+-}_{\bl=0}(\Om)/\hbar$  by comparing the rate equation $d\langle N(t)\rangle/dt=-(\gamma_{\downarrow}-\gamma_{\uparrow})\langle N(t)\rangle +\gamma_{\uparrow}$ for the mean occupation $\langle N(t)\rangle\equiv\sum_{n=0}^{\infty}nP_{n}(t)$ with the phonon-energy balance equation of Refs.~\onlinecite{Paulsson:RapCom,Frederiksen:PRB07}. The stationary state has a geometric distribution $P_{n}(t\to\infty)\propto(\gamma_{\uparrow}/\gamma_{\downarrow})^{n}$ with the correct asymptotic mean $\langle N(t\to \infty)\rangle=(\gamma_{\downarrow}/\gamma_{\uparrow}-1)^{-1}=\bar{N}$ and the exponentially decaying connected correlation function $\langle\!\langle N(t)N(0)\rangle\!\rangle\equiv \langle N(t)N(0)\rangle-\langle N(t)\rangle\langle N(0)\rangle=\langle\!\langle N^{2}\rangle\!\rangle e^{-|t|/\tau_{\mathrm{rel}}}$, with the inverse relaxation time $\tau_{\rm rel}^{-1}=\gamma_{\downarrow}-\gamma_{\uparrow}=2|\mathrm{Im}\Pi^{R}(\Om)|/\hbar=2\gamma_{e\rm ph}\Om/\hbar\pi$ and variance $\langle\!\langle N^{2}\rangle\!\rangle=\bar{N}(\bar{N}+1)$. At large voltage the correction to the mean current due to $e$-ph coupling is given by $I_{\mathrm{eph}}(eV\gg\Om)\approx \tfrac{e\gamma_{e\mathrm{ph}}}{\pi\hbar}(\bar{N}+\tfrac{1}{2})(3-4\mathcal{T})eV\equiv i[\bar{N}]$.\cite{Haupt} Since the relaxation time $\tau_{\rm rel}\sim \hbar/(\gamma_{e\mathrm{ph}}\Om)\gg \hbar/\Om\gg\hbar/eV$ is the longest time scale in the problem, the noise at large voltage can be estimated assuming the current to follow adiabatically the phonon occupation $N(t)$,   
$S_{\rm eph}(eV\gg\Om)=\int_{-\infty}^{\infty} dt \langle\!\langle i[N(t)] i[N(0)]\rangle\!\rangle= \left(\tfrac{e\gamma_{e\mathrm{ph}}}{\pi\hbar}\right)^{2}(3-4\mathcal{T})^{2}(eV)^{2}\int_{-\infty}^{\infty} dt \langle\!\langle N(t)N(0)\rangle\!\rangle$. Putting everything together, we arrive at the microscopic result of Eq.~\eqref{Sasym}. 

The same line of reasoning can be repeated for a general multilevel junction without the need of eWBA. We can thus generally conclude that at large voltage (i)  the correction to noise due to $e$-ph coupling is dominated by the backaction term $S_{\rm ba}$ and (ii) this term is directly related to the diffusion of the energy stored in the oscillator, which fluctuates slowly because of the randomness of the driving tunneling events.  The typical frequency of these fluctuation $\tau_{\rm rel}^{-1}$ is well above standard schemes for filtering out $1/f$-components  ($\tau_{\rm rel}^{-1}\sim 100 \rm GHz$ for $\hbar\Om\sim10 {\rm meV},\gamma_{e\rm ph}\sim 0.01$), so that $S_{\rm ba}$ will play a relevant role in noise measurements in systems with weak phonon damping. 

The backaction term $S_{\rm ba}$ has also important consequences around the phonon emission threshold $eV=\Om$. 
In particular, it affects the inelastic noise signal $\Delta S'=\left.\tfrac{\partial S}{\partial V}\right|_{eV=\Om^{+}}-\left.\tfrac{\partial S}{\partial V}\right|_{eV=\Om^{-}}$, which is a  quantity of prime experimental interest defining the behavior of noise around the phonon emission threshold at low temperature. As an example, in the considered case of symmetrically coupled single-level junctions (in which $\Delta S'$ can be simply expressed in terms of the transmission of the junction) the full result for nonequilibrated phonons  $S_{\rm av}+S_{\rm ba}$ leads to  $\Delta S'\propto24\mathcal{T}^{2}-30\mathcal{T}+17/2$, which changes sign at transmissions $\mathcal{T}_{1,2}\doteq 0.434$ and $0.816$. This should be contrasted with the values $\mathcal{T}_{1,2}\doteq 0.146$ and $0.854$,\cite{Schmidt} valid for completely thermalized phonon.\footnote{Phonon heating also affects the crossover between enhancement and reduction of the conductance due to $e$-ph coupling. For the single-level model this crossover shifts from $\mathcal{T}=1/2$ to $\mathcal{T}=5/8$ in the two opposite limits of strong and negligible external phonon damping. }

Finally, we have applied the eWBA multilevel results given in the supplemental material to the case of an atomic gold wire with unitary transmission and no external phonon damping,\cite{Note3} as shown in Fig.~\ref{fig:SAuchain}. Remarkably, including $S_{\rm ba}$ more than doubles the noise above the phonon emission threshold as compared to the ``static'' contribution $S_{\rm av}$ alone. For comparison we also plotted the noise in the case of fully thermalized phonons.  An intermediate level of equilibration simply interpolates between the two limits.\cite{Note3} This example clearly shows the need for taking into account the phonon backaction correction in order to properly describe the noise above the phonon emission threshold.
\begin{figure}
\begin{center}
{\resizebox{0.9\columnwidth}{!}{\includegraphics{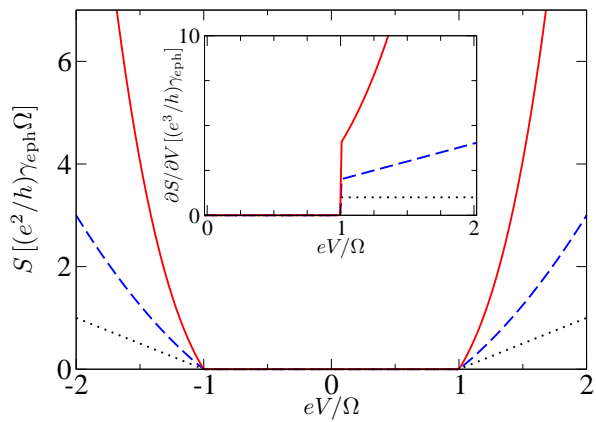}}}
\end{center}
\caption{(Color online) Zero-temperature current noise in a perfectly transmitting atomic gold wire in the presence of an undamped alternating-bond-length phonon mode.\cite{Frederiksen:PRL04,Paulsson:RapCom} In this case $S_{\rm elastic}=0$. Solid red line, $S=S_{\rm av}+S_{\rm ba}$;  dashed blue line, $S_{\rm av}$; dotted black line, case of thermally equilibrated phonons; inset, voltage derivative of the noise.}\label{fig:SAuchain}
\end{figure}

In conclusion, we have evaluated the full lowest-order expression for the current noise in the presence of an externally undamped phonon mode, identifying a purely nonequilibrium  contribution due to the backaction of the heated phonon. At high voltage, this contribution dominates the noise and it can be fully understood in terms of slow fluctuations of the phonon occupation. Importantly, this nonequilibrium correction also modifies the quantum behavior of the noise around the phonon emission threshold and has, therefore, direct implications for ongoing  experiments.

We thank D.~Bagrets, J.~M.~van~Ruitenbeek, and Yu.~V.~Nazarov for
useful discussions. We acknowledge financial support by the Czech Science Foundation via Grant No.~205/10/0989, the Ministry of Education of the Czech
Republic via the research plan MSM 0021620834, the Visiting Professors Program of  Universit{\'e} Joseph Fourier in Grenoble (T.~N.), and the DFG via
SFB 767. \\ T.~N. and F.~H. contributed equally to this work.

%


\end{document}